\documentclass[useAMS,usenatbib]{mn2e}
\usepackage[dvips]{graphicx}
\usepackage{amsmath}

\title[Lightcurve analysis for eccentric orbits]{Transiting planets - lightcurve analysis for eccentric orbits}
\author[David M. Kipping]{David M. Kipping$^{1}$\thanks{E-mail:
d.kipping@ucl.ac.uk}\footnotemark[1]\\
$^{1}$Department of Physics and Astronomy, University College London, \\
       Gower Street, London WC1E 6BT, UK}
\begin{document}

\date{Accepted 2008 June 27. Received 2008 June 26 ; in original form 2008 March 4}

\volume{389} \pagerange{1383--1390} \pubyear{2008}

\maketitle

\label{firstpage}

\begin{abstract}
Transiting planet lightcurves have historically been used predominantly for measuring the depth and hence ratio of the planet-star radii, $p$. Equations have previously been presented by  \citet{sea03} for the analysis of the total and trough transit lightcurve times to derive the ratio of semi-major axis to stellar radius, $a/R_*$, in the case of circular orbits.  Here, a new analytic model is proposed which operates for the more general case of an eccentric orbit.  We aim to investigate three major effects our model predicts: i) the degeneracy in transit lightcurve solutions for eccentricity, $e>0$ ii) the asymmetry of the lightcurve and the resulting shift in the mid-transit time, $T_{MID}$ iii) the effect of eccentricity on the ingress and egress slopes.  It is also shown that a system with changing eccentricity and inclination may produce a long period transit time variation (${LTTV}$).  Furthermore, we use our model in a reanalysis of HD209458b archived data by \citet{ric06}, where we include the confirmed non-zero eccentricity and derive a $24 \mu m$ planetary radius of  $R_P=1.275R_J \pm 0.082 R_J$ (where $R_J$ = 1 Jovian radius), which is $\sim 1\%$ larger than if we assume a circular orbit.
\end{abstract}


\section{Introduction}

In 2000, the first transiting extra-solar planet was detected by \citet{cha00} and \citet{hen00} using the increasingly prosperous occultation method. \citet{sea03} proposed that there are four principle pieces of information retrievable from any given lightcurve, being transit depth, $\Delta F$, total transit duration, $t_T$, transit duration between ingress and egress, $t_F$, and mid-transit time, $T_{MID}$ (for which consecutive measurements may determine the orbital period).  The Seager \& Mall\'{e}n-Ornelas (SMO) equations were constructed under the assumptions of a circular orbit and no limb darkening.  However, as more and more eccentric transiting planets are discovered, the need for a more general set of equations has become increasingly exigent.

From the 33 transiting planets \emph{with measured eccentricities}, over 25\% have eccentric orbits\footnote{See http://exoplanet.eu by J. Schneider} and in such cases the SMO equations will no longer be valid.  In addition, transit detection probabilities favour eccentric systems by a factor of $(1-e^2)^{-1}$, as predicted by \citet{bar07}.  This information implies that lightcurves measured for eccentric transiting planets will become common and thus a quick, efficient and accurate way of analyzing them is necessary.  Employing circularized equations to eccentric planets will clearly lead to unnecessary systematic errors of the system parameters, as pointed out by \citet{bar07}.

Our intention is to formulate a new model which improves upon that by \citet{sea03} by allowing for orbital eccentricity.  Hence, the new model must analytically reduce to the SMO equations in the case of $e=0$.  These equations will be able to produce sample lightcurves for a given set of system parameters which can then be compared to the real lightcurve to find a best fit.  

It will also be shown that eccentricity cannot be determined from the primary transit lightcurve \emph{alone} with current telescopes and any single lightcurve may be generated by a variety of input parameters.  Another motivation for this work is to produce a set of analytic equations rather than present a computational technique for such a complex problem.  This will allow observers to produce sample lightcurves very quickly for any chosen system parameters and reduce post-observation processing time.

We note that other authors have previously discussed the effects of eccentricity on transiting planets.  \citet{brk07} and \citet{bar07} independently focussed their discussions onto the probability of detection of eccentric planets, something that we do not consider in this work.  In contrast, \citet{tin05} and \citet{for08} presented models for the analysis of eccentric lightcurves.  This work is a more elaborate development of the model developed these latter authors, which we compare our model to in detail in \S5.1.

In \S2, we briefly describe the model and the underlying assumptions (a detailed derivation may be found in the appendix).  In \S3, we discuss the implications of our model and degenerate solutions to a lightcurve.  In \S4, we show some typical simulations of the model for a typical system and in \S5.1 compare its performance with the models by \citet{tin05} and \citet{for08}.  In section \S5.2, we discuss lightcurve asymmetry and a new effect where changing orbital parameters can cause a long term transit time variation (${LTTV}$).  In section \S5.3, we employ our model in a reanalysis of the HD209458b data taken by \citet{ric06}, where we derive a new $24 \mu m$ planetary radius of  $R_P=1.275R_J \pm 0.082 R_J$, which is $\sim 1\%$ larger than assuming a circular orbit.

In section \S5.4, we discuss the effect eccentricity has on the ingress and egress slope shapes and conclude present-day telescopes cannot reliably determine eccentricity from lightcurve data \emph{alone}.  Finally, in section \S6 we discuss our findings and conclusions.

\section{The Model}

Two principle inputs for the model we present here are the total and trough duration times ($t_T$ and $t_F$ respectively), which are independent of limb darkening effects.  Although, we stress here that the model can be coupled with the equations and code written by \citet{man02} to produce limb darkening corrected lightcurves (see the appendix).  The definitions used in this paper are the same as that used by \citet{sea03} and may be found in their paper, figure 1.  Throughout, we make the following assumptions:

\begin{itemize}
\item The light comes from a single star, rather than from two or more blended stars.
\item The target star and planet are perfect spheres with definite edges.
\item The companion is dark.
\item The orbital period of the planet is known (from radial velocity surveys or consecutive $T_{MID}$ times).
\item The mass of the planet (or radial velocity semi-amplitude) is approximately known\footnote{Approximate value acceptable since $M_*$ dominates.}.
\end{itemize}

Note, that we do \emph{not} make the following assumptions, which have been adopted in previous models:

\begin{itemize}
\item The planetary orbit is circular, \citet{sea03}.
\item $M_P \ll M_*$ where $M_P$ is the mass of the planet, $M_*$ is the mass of the star, \citet{sea03}.
\item The mean planet-star separation is much greater than both the stellar radius, $R_S$, and the planetary radius, $R_P$, \citet{for08}.
\item The transit duration is much less than the orbital period, $P$, \citet{for08}.
\item The planet-star separation during the transit is nearly constant, \citet{tin05} and \citet{for08}.

\end{itemize}

A transit starts when the outer rim of the planet makes contact with the outer rim of the star, as seen from the Earth.  This is defined as contact point $1$ in the forthcoming analysis.  In a similar way we define contact point $4$ to be the exit point.  We note that both points have a projected planet-star separation of $R_*+R_P$ ($L_B$), where $R_*$ \& $R_P$ are the stellar \& planetary radii respectively.  We define contact point $2$ as the point when the projected planet is now fully inside projected star, and similarly point $3$ for the exit (figure 1 of \citet{sea03}).  These latter two positions correspond to the start at end of the trough section of the lightcurve and have a projected planet-star separation of $R_*-R_P$ ($L_S$).

Our model allows for changing angular velocity as a function of true anomaly, as well as changing planet-star separation, unlike many previous analytic models.  We found that the most accurate way to model such a system is to construct a full 3D geometric formulation and employ the conservation of angular momentum.  In doing so, an elegant analytic expression is no longer possible and there is an argument for a numerical approach.  However, in this model, we do not adopt such a methodology, since analytic arguments offer lower computation times and the ability to reduce to special cases (such as $e=0$, as shown in the appendix).

In our model, we consider an elliptical orbit in a standard frame, $S$, and rotate it for position of pericentre and orbital inclination into a new frame, $S_{FINAL}$.  This new frame represents what an observer from the Earth would be able to see, with the star centred on the origin (i.e. the projected orbit).  We then find the intersection points of the projected orbit with a circle of radius $L_B$ and $L_S$ respectively, which give us the full range of transit contact points.  We are then able to back-transform and derive the contact point true anomalies, which we label as $f_1$, $f_2$, $f_3$ and $f_4$.  For details of the derivation, please see the appendix.  We then convert the angles into times using conservation of angular momentum (\citet{ser02}).  One significant advantage of the model is that we derive the projected star-planet separation, which can be used as a direct input into the equations and code of \citet{man02}, allowing for limb darkening correction.  Using the function $D(f)$, which we call the \emph{duration function}, we find that the time between any two contact points $a$ and $b$ is given by:

\begin{equation}
t_a - t_b = \frac{\mu a^2}{J} [D(f_a) - D(f_b)]
\end{equation}
where $D(f)$ is defined in the appendix, $J$ is the planet's angular momentum, $\mu$ is the reduced mass and $a$ is the orbital semi-major axis

\section{Implications}

Any single lightcurve has three easily distinguishable characteristics; $p$ (ratio of planet-star radii), $t_T$ and $t_F$.  Transits allow the measurement of orbital period, $P$, as well through the time difference between consecutive transits.  For a hypothetical planet-star system, all four of these values may be derived using Kepler's laws and the following inputs: 

\begin{enumerate}
\item Radius of the star, $R_*$
\item Radius of the planet, $R_P$
\item The mass of the star $M_*$ (which determines $a$)
\item The mass of the planet, $M_P$ (or alternatively the radial velocity semi-amplitude, $K$)
\item The orbital inclination, $i$
\item The eccentricity, $e$
\item The position of pericentre, $\varpi$
\end{enumerate}

It therefore becomes clear that seven variables produce the resulting lightcurve shape and so there will be a degeneracy in the four distinguishing features of the lightcurve, namely $p$, $t_T$, $t_F$ and $P$.  In other words, different combinations of system parameters can produce effectively identical lightcurves.  The implication from this is that eccentricity, $e$ and $\varpi$ cannot be determined from primary lightcurve data \emph{alone}.  Additional information is required.  The model we present here can produce lightcurves for any given system parameters, but we emphasize that it is possible to produce ostensibly identical lightcurves from different input parameters.

\citet{sea03} appreciated that if an observer wishes to avoid using stellar mass-radius relationships, then one can only determine the $R_*$ (and hence $R_P$) from a lightcurve if one has an independent and accurate measurement of the stellar mass.  Once $M_*$ is known, $R_*$ can be derived from the lightcurve data alone.  We now extend this line of thought to include $e$ and $\varpi$.  Both of these parameters must \emph{also} be determined by an independent observation, along with stellar mass, before the radius of the star can be determined.  

The ideal observation procedure would therefore be to use an independent measurement of $M_*$, followed by radial velocity measurements to determine $e$, $\varpi$ and $M_P$ and then finally use transit observations to calculate $P$, $R_*$ and finally $R_P$.  In the absence of an independent $M_*$ measurement, stellar evolution must be invoked.  In section \S5.3, we will apply a typical analysis to HD209458b, but we first compare our model with two other established models.

\section{Validation of the Model}

For our tests of the model, we consider a Solar star, with a Jupiter planet at a 3-day period orbit (so $a = 0.0407 {AU}$).  Consider that this planet to have a position of pericentre of $20^\circ$.  We now consider a range of eccentricities and produce the transit time durations for different inclinations.  Figures 1 \& 2 show the results for multiple cases for different eccentricities, positions of pericentre and inclinations.

Any new model proposed, must be able to be demonstrate that the SMO equations are a special case of the general form, for $e=0$.  For a circular orbit, we use the exact same assumptions as the SMO equations and therefore an important test of this new model is that it should reduce down to that form.  Running a wide range of numerical calculations, we find that our model agrees precisely with that of the SMO equations for both $t_T$ and $t_F$ for circular orbits\footnote{Note that $t_T$ and $t_F$ are independent of limb darkening}. Indeed, we show in the appendix that the equations are analytically identical in such conditions.

\section{Results}
\subsection{Comparison \& Tests of the Model}

We find our model disagrees with the model by \citet{tin05} and \citet{for08}.  We note that the models by these two authors are analytically identical and therefore there is no need for us to compare to both.  In figure 3, we see that there is a systematic discrepancy of around 30 seconds between Ford et al.'s model and that of both the model we present here and the SMO model for $e=0$\footnote{For the case of the typical system we have described}.  This systematic persists up to $e\approx0.5$ and then diverges up to several minutes for very high $e$.  We attribute the source of discrepancy to the different analytic expressions for the case of $e=0$, where we observe that the model we present here reduces down to the SMO equations whereas that of Tingley \& Sackett and Ford et al. do not, but are, however, simpler and quicker to utilize.

\begin{figure}
\begin{center}
\includegraphics[width=8.4 cm]{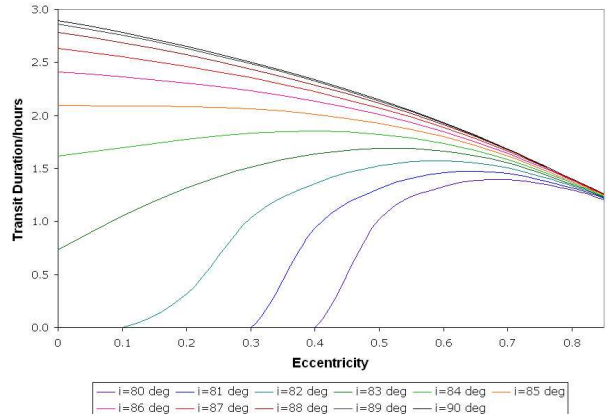}
\caption{\emph{Calculations for $t_T$ using our model for various $e$ and $i$.  A convergence occurs for very high $e$, making an accurate determination of $i$ more difficult.  Note how closely spaced the high inclinations are, making differentiation difficult here too.  Also, we note that our model cannot generate the durations for orbits of $e>0.885$ since the perihelion is inside the star's radius for such orbits.}} \label{fig:fig1}
 \end{center}
\end{figure}

\begin{figure}
\begin{center}
\includegraphics[width=8.4 cm]{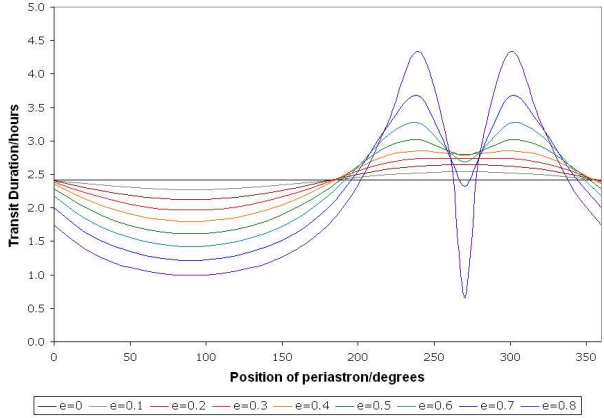}
\caption{\emph{Variation of $t_T$ for a typical system as a function of $\varpi$ and $e$.  We place a Jupiter with a 3-day orbit around a Solar star at $i=86^{\circ}$.  The observed sinusoidal shape is due to the change in velocity at different true anomalies and the sudden dip at $\varpi \sim 270^\circ$ is due to the impact parameter approaching $(1-p)$ due to increasing planet-star separation.}} \label{fig:fig2}
 \end{center}
\end{figure}

We find that for very high eccentricities, the transit time duration varies very little between different inclinations, making an accurate determination of $i$ problematic.  In this case, we have chosen $\varpi=20^\circ$, but the same effect occurs across a range of values for $\varpi$.  The source of this is due to our test planet being in a very close orbit and so when we have high eccentricity, it grazes the star's surface and inclination becomes progressively less important.  The effect is therefore typical but not general.

We also observe how near $\varpi=270^\circ$, there is a dip in the transit time duration (figure 2) due to the impact parameter rapidly increasing as the planet enters aphelion, increasing the planet-star separation to a maximum; an effect which we expect to see.

\begin{figure}
\begin{center}
\includegraphics[width=8.4 cm]{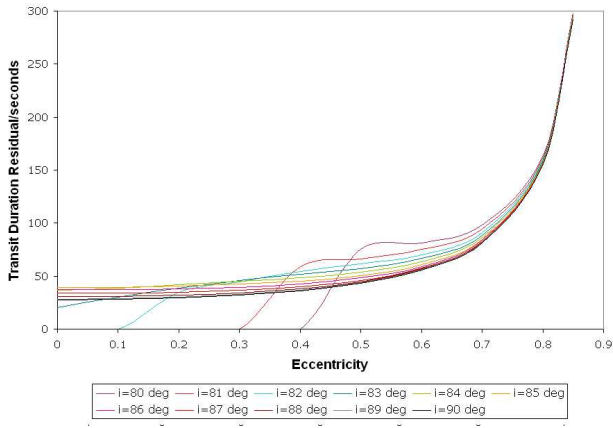}
\caption{\emph{Residuals in $t_T$ between the model we present here and Ford et al.Õs model, for a typical system.   The difference is around 30 seconds until we reach high $e$ where it diverges to minutes.}} \label{fig:fig3}
 \end{center}
\end{figure}

\subsection{Lightcurve Asymmetry \& $T_{MID}$ Shifts}

Aside from solution degeneracy, the other major effect of eccentricity on a lightcurve is asymmetry, as discussed recently by  \citet{bar07}.  Since velocity is a function of phase, we can see that the ingress and egress slopes need not be of equal duration and that mid-transit time, $T_{MID}$, will be shifted away from the halfway point between ingress and egress.  The direction of the shift will be a function of position of pericentre only, and not the sense of orbital rotation.  The principal motivation for such a detailed investigation is that $T_{MID}$ is predicted to be a powerful future tool for exomoons detection (see \citet{sim07}) and perturbing bodies (see \citet{hol04}).  We define $T_{MID}$ shift as the difference between the apparent mid-point and the true mid-point.  Below, we present our mathematical definition and in figure 4 we show a range of shifts for different system parameters in a typical system.

\begin{equation}
T_{MID,apparent} =t_{REF} +  \frac{\mu a^2}{J} \frac{[D(f_4)-D(f_1)]}{2}
\end{equation}
\begin{equation}
T_{MID,true} =t_{REF} +  \frac{\mu a^2}{J} [D(f_{AV}) - D(f_1)]
\end{equation}
\begin{equation}
\Delta T_{MID} =T_{MID,true}-T_{MID,apparent}
\end{equation}
\emph{where $f_{AV}=\pi/2-\varpi$}

\begin{figure}
\begin{center}
\includegraphics[width=8.4 cm]{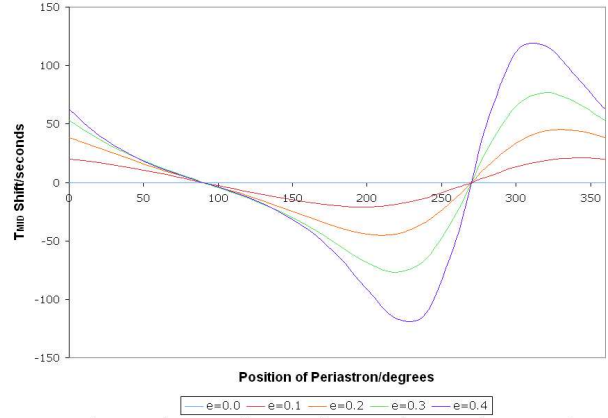}
\caption{\emph{The predicted shift in $T_{MID}$ for different eccentricities and $\varpi$, for a typical 3-day period system.  The effect on $T_{MID}$ becomes quite dramatic for very eccentric systems near $230^\circ$ and $310^\circ$.}} \label{fig:fig4}
 \end{center}
\end{figure}

Transit time variation (${TTV}$) requires $T_{MID}$ to be accurate to the second level for detections of perturbing bodies (see \citet{sim07}).  If the target system were stable, then compensating for this shift would not be necessary, since it would only cause a constant systematic error which would not affect the variation of $T_{MID}$ about some average.  However, if we have a perturbing body, an eccentric transiting planet can expect to experience changing $e$ or $\varpi$ over time.  The result is that $\Delta T_{MID}$ will have changed too over such a time-scale.

Ergo, a perturbing body will cause two ${TTV}$ signals, one short period signal due to the effect described by \citet{ago05} ($TTV$), and one long period signal ($LTTV$) due to a combination of the expected secular changes and our $T_{MID}$ measurement method.  The difference between the two is subtle.  The short period signal is a real effect, whereas the effect we describe here is a consequence of the measurement technique and hence a systematic error effect.  Secular variations can cause $TTV$, but here the secular variations create an additional long period signal simply due to the fact we typically define the mid-transit time as being halfway between the ingress and egress.  If we always defined the mid-transit time as being the moment when the planet crosses the star's central line ($T_{MID,true}$), then this signal would not be present.

In the case of the proposed, but retracted, planet GJ436c,  \citet{rib08} suggested an important test would be the variation of $T_{MID}$ of GJ436b, but this planet has a significant eccentricity of 0.15, as measured by \citet{dem07}.  For this planet, we predict $\Delta T_{MID}$ is currently 20.95 seconds.  If $e$, $i$ or $\varpi$ are changing over time (predicted if gravitational interactions with a third body occur by Ribas et al.), then $\Delta T_{MID}$ will also change (see figure 5).

In figure 4, we illustrate the shift in $T_{MID}$ for the 3-day Jupiter test system we discussed in \S4, for different $e$ and $\varpi$.  In figure 5, we produce the long period ${TTV}$ signal for the case of GJ436b due to the $T_{MID}$ effect alone, where $e$ is changing sinuoidally over a 30 year timescale with an amplitude of 0.1 (values proposed by Ribas in personal communication).  We keep position of pericentre fixed at $351^\circ$ (\citet{man07}) and allow $i$ to change in accordance with that predicted by \citet{koz62} to conserve angular momentum.  If $\varpi$ is increasing over time, the $T_{MID}$ signal would decrease and vice versa.  Similarly, if inclination modulates with a larger amplitude, the $T_{MID}$ signal would increase too and vice versa.

\begin{figure}
\begin{center}
\includegraphics[width=8.4 cm]{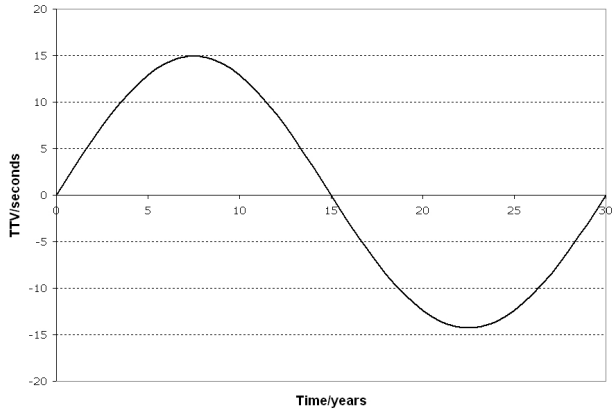}
\caption{\emph{An example of a possible LTTV signal due to changing eccentricity and inclination for GJ436b.  Such changes could be induced by a perturbing body.  The effect is a result of our measurement technique plus secular variations.  Over a period of a few years, the LTTV signal could manifest itself as transits being progressively late or early.}} \label{fig:fig5}
 \end{center}
\end{figure}

As well as $T_{MID}$ being shifted, the ingress and egress slopes can differ in duration for eccentric systems.  Similar simulations showed that the difference between the ingress and egress slopes is on the order of sub-seconds to tens of seconds for very eccentric systems close to $\varpi=0^o$ and $\varpi=180^o$.  The effect is unlikely to prove useful, in general, for diagnosing $e$ or $\varpi$ due to typical timing errors.  For example, in the case of the highly eccentric system HD17156b ($e=0.6717$ as measured by \citet{gil07}), the difference in slope times is predicted to be just 0.4 seconds.

\subsection{A Case Study for HD209458b}

Ever since its discovery, HD209458b has been a topic of debate in regard to eccentricity, or lack there of (\citet{lau05}, \citet{win05}, \citet{knu07}).  In reality, no orbit can be truly circular, and indeed a study by \citet{win05} showed that $0.0057<e<0.023$ to within 90\% confidence level.  \citet{lau05} came to a similar conclusion.  However, many previous analyses of HD209458b have employed the SMO equations (e.g. \citet{tor08}), which means the analyses will be somewhat flawed in assuming $e=0$, when we know that this is not true, as determined by \citet{win05}.  

We were interested in investigating how much difference including such a small eccentricity would have on the final result.  To our knowledge, no previous authors have included eccentricity in the lightcurve analysis of HD209458b either, making such a study highly valuable.  On a side note, we stress here a subtle point; that we do not, and indeed in practice \emph{cannot}, use the lightcurve data alone to determine orbital eccentricity (see \S5.4).

For this case analysis, we used the $24 \mu m$ Spitzer result from \citet{ric06} because the lightcurve is not significantly affected by limb darkening and thus Richardson et al. were able to derive precise timing values of $t_T=2.979 \pm 0.051$ hours and $t_F = 2.253 \pm 0.058$ hours and a transit depth of $\Delta F=0.01493 \pm 0.00029$ (see table 1 for full list).

Our method was to solve our equations for $R_*$ and $i$ across a range of values for $M_*$, from $0.9 M_{\bigodot}$ to $1.3 M_{\bigodot}$ in $0.5 M_{\bigodot}$ steps and interpolate in-between these points.  The values and errors of the other parameters are all known and so in this way we may generate a lightcurve derived mass-radius relationship for the star (L-MRR).  We solve our equations by employing a nested secant iteration of our equations, using the model by \citet{for08} as a first guess.  Once the optimal solution has been found, we turn our attention to calculating the error bar on the L-MRR.

In this work, we derive our errors by employing a Monte Carlo bootstrap procedure with 1000 steps for selected values of $M_*$.  This is done at each step value of $M_*$ and then we once again interpolate in-between the steps to generate a continuous function.  In each Monte Carlo run, $t_T$, $t_F$, $p$ and $e$ are randomly allocated according to a Gaussian probability distribution with the mean located at their best-known value and a standard deviation set to be the quoted one-sigma errors.  The selected values and errors are shown in table 1.  We do not consider the error in the radial velocity semi-amplitude, $K$, the orbital period, $P$, or the argument of periastron, $\varpi$, due to their very small effect on our final result and to minimize computation time.  After this stage, we have now produced the L-MRR and one-sigma confidence limits, which can be seen in figure 6.  We then repeat the whole process but using the SMO equations (i.e. assuming $e=0$) as a comparison.

We use the eccentricity derived by \citet{lau05} and \citet{win05} of $e = 0.014 \pm 0.005$.  Since the eccentricity is small, the effect of $\varpi$ is minimal and hence we choose to fix it to the best known-value of $\varpi=83^\circ$ as calculated by \citet{win05}.  We also fix the orbital period, since \citet{knu07} were able to derive a very precise value of $P = (3.52474859 \pm 3.8\times10^{-7})$ days.  Finally, we fix the radial velocity semi-major amplitude to that derived by \citet{wit05} of $K =  82.7 \textrm{ms}^{-1}$, which is justified because both the error and the effect of the parameter are small on our results.  So the confidence limits on our L-MRR are derived from the errors on $t_T$, $t_F$, $p$ and $e$.

To find the unique solution in this case, one must employ stellar evolution, since the errors bars on the spectropsically determined mass are quite large (\citet{maz00}).  \citet{cod02} provide a stellar evolution MRR by fitting stellar models to constant luminosity.  This MRR has a negative gradient and so intersects our own L-MRR.  We find the intersection point and propagate our errors to derive our final result.  The inputs and outputs of the whole procedure are summarized in table 1.

\begin{figure}
\begin{center}
\includegraphics[width=8.4 cm]{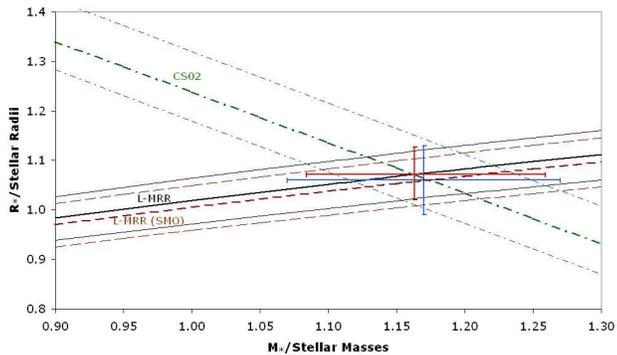}
\caption{\emph{Lightcurve derived mass-radius relation (L-MRR) for HD209458.  The line labeled L-MRR (SMO) is the same relation derived assuming $e=0$.  The line labeled CS02 is the stellar evolution relation derived by \citet{cod02}.  Fainter lines indicate one-sigma confidence limits.  We overlay two data points, the lower one being that of \citet{ric06} (blue) and the slightly higher one (red) being that of this work.}} \label{fig:fig6}
 \end{center}
\end{figure}

\begin{table*}
\caption{\emph{Summary of bootstrap inputs and results.  Columns 2 \& 3 are the derived results using a bootstrap method and stellar evolution.  K08 refers to the model presented in this paper, SMO refers to the model by \citet{sea03}.  The final column shows the quoted results by \citet{ric06} (R06), where we display the values as they appear in the publication.}} % title of Table
\centering % used for centering table
\begin{tabular}{c c c c} % centered columns (2 columns)
\hline\hline %inserts double horizontal lines
Parameter & K08 & SMO & R06 \\ [0.5ex] % inserts table
%heading
\hline % inserts single horizontal line
\emph{Inputs} \\
\hline
$P$/days & 3.52474859 & 3.52474859 & 3.52474859 \\
$K/\textrm{ms}^{-1}$ & $82.7 \pm 1.3$ & $82.7 \pm 1.3$ & - \\
$e$ & $0.014 \pm 0.005$ & 0 & 0 \\ 
$\varpi$/$^{\circ}$ & 83 & - & - \\
$t_T$/hours & $2.979 \pm 0.051$ & $2.979 \pm 0.051$ & $2.979 \pm 0.051$ \\
$t_F$/hours & $2.253 \pm 0.058$ & $2.253 \pm 0.058$ & $2.253 \pm 0.058$ \\
$(R_P/R_*)$ & $0.1222 \pm 0.0012$ & $0.1222 \pm 0.0012$ & $0.1222 \pm 0.0012$ \\
\hline
\emph{Results} \\
\hline
$M_*$/$M_{\bigodot}$ & $1.163_{-0.079}^{+0.096}$ & $1.174_{-0.079}^{+0.096}$ & $1.171$\\
$R_*$/$R_{\bigodot}$ & $1.072_{-0.052}^{+0.055}$ & $1.060_{-0.051}^{+0.054}$ & $1.064 \pm 0.069$ \\
$\rho_*$/$(\textrm{kg m}^{-3})$ & $1330 \pm 230$ & $1390 \pm 240$ & - \\
$i$/$^{\circ}$ & $87.94 \pm 0.76$ & $88.00 \pm 0.76$ & $88.00 \pm 0.85$ \\
$M_P$/$M_J$ & $0.681 \pm 0.039$ & $0.690 \pm 0.039$ & - \\
$R_P$/$R_J$ & $1.275 \pm 0.082 R_J$ & $1.261 \pm 0.081$ & $1.265 \pm 0.085$ \\
$\rho_P$/$(\textrm{kg m}^{-3})$ & $407 \pm 81$ & $426 \pm 85$ & - \\ [1ex] % [1ex] adds vertical space
\hline %inserts single line
\end{tabular}
\label{table:nonlin} % is used to refer this table in the text
\end{table*}

We derive the 24$\mu m$ planetary radius to be $R_P=1.275R_J \pm 0.082 R_J$.  The result derived using the SMO equations is almost identical to that of \citet{ric06} (see table 1) and we attribute the slight differences in the values due to the fact we use a fixed value for $K$ rather than a fixed value for $M_*$.  

It is found that negating eccentricity results in an systematic underestimation of the stellar radius of HD209458 at the 1\% level, even with a very low value for $e$.  We believe this outlines the importance of incorporating eccentricity into lightcurve analyses of other planets, even those of low $e$.

\subsection{Ingress \& Egress Slopes as a Function of Eccentricity}

Even in the absence of limb darkening, the ingress and egress slopes of the lightcurve are not straight lines but curves, since $\frac{\textrm{d}A}{\textrm{d}t} \neq constant$ during the slope part of the transit (where $A$ is the occulting area).  It may easily shown, that for any orbit, the occultation of the star will have an S-shape (\citet{man02}).  We now take these equations and apply our new eccentric orbit model to produce the slopes.  This will allow us to investigate the effect of eccentricity on such curves, which, to our knowledge, has not been investigated in detail.

Consider the typical 3-day period system once more.  If we set $\varpi=90^\circ$ then the ingress and egress slope times will be identical and we have a symmetric lightcurve.  We then choose $i=86^\circ$ and calculate the slope shapes for different eccentricities.  All the slopes will have the same start and end depth but different durations.  In order to compare the shapes only, we scale the slope duration to go from 0 to 1 in each case, in a unit we label \emph{relative time}.  

Now let us consider whether it is possible to determine the eccentricity simply from the ingress curvature.  In figure 7, we show the residuals between the ingress curvatures for different eccentricities in comparison to an $e=0$ shaped line.  We find that $\Delta e \approx 0.1$ gives a $\sim 5 \times 10^{-5}$ change in normalized flux, which would be beyond the sensitivity of current telescopes (e.g. \citet{bea08}) but potentially feasible with the JWST mission\footnote{See http://www.jwst.nasa.gov/}.  However, such an observation is highly unlikely to be able to better radial velocity measurements of $e$.

\begin{figure}
\begin{center}
\includegraphics[width=8.4 cm]{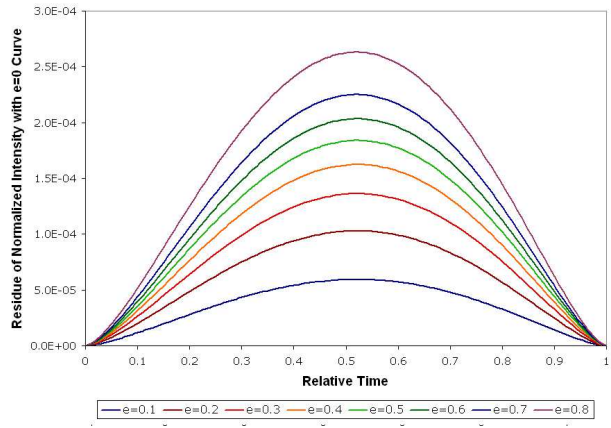}
\caption{\emph{Residuals of ingress curvature of different eccentricities compared to the $e=0$ curve.  A difference of 0.1 eccentricity would be typically indistinguishable with current telescopes, but probably not with the JWST telescope.}} \label{fig:fig7}
 \end{center}
\end{figure}

We conclude that although it is unfeasible for current telescopes to distinguish between different eccentricities using the ingress/egress slope shape or the difference between ingress and egress durations, future missions could do so.  Ideally, these measurements should in the infrared, where limb darkening would be less significant.

\section{Discussion and Conclusions}

We have presented a new model for calculating the transiting planet lightcurves, which uses a 3D geometrical interpretation of planetary systems and may be integrated with the Mandel-Agol code to produce highly accurate lightcurve simulations.  We have shown that the Seager \& Mall\`en-Ornelas equations are a special case of our more general formulation, for when the eccentricity is zero.

The model we present here requires two additional pieces of information than is needed for circular orbit lightcurve analysis; namely eccentricity and position of pericentre.  Using archived data and including the confirmed non-zero eccentricity of HD209458b, we derive a revised $24 \mu m$ planetary radius of $1.275 \pm 0.082 R_J$, which is 1\% larger than if we assume $e=0$.  We propose that the ideal method for determining a stellar and hence planetary radius is to use measurements independent of the primary transit to determine $M_*$, $e$, $\varpi$ and $M_P$ and then use the transits to determine $P$, $R_*$ and finally $R_P$.  In the case of no independent $M_*$ measurements, stellar evolution must be invoked.

We also showed that the mid-transit time would be shifted off-centre due to lightcurve asymmetry (for eccentric orbits) from seconds to minutes depending upon $e$.  Furthermore, changing system parameters, perhaps due to a perturber, could produce a long period $TTV$ signal ($LTTV$) of the order of tens of seconds.  This could manifest itself as progressively late or early transits over a few years time-scale.

Finally, we have included an analysis on the egress and ingress slope curvature as a function of eccentricity and concluded that current telescopes will be unable to distinguish the eccentricity or position of pericentre from lightcurve data alone.  We propose that JWST could be able to make such a measurement but is unlikely to match the precision produced by radial velocity surveys.

\section*{Acknowledgments}

D. M. K. has been supported by UCL and the Science Technology and Facilities Council (STFC) studentships.  The author would like to thank Jeremy Richardson for taking such a precise set of data for HD209458b, which proved to be an ideal test bed for the new model.  The author would also like to thank Ignasi Ribas for his time in technical discussions regarding the model and for providing estimates for the evolution of GJ436b, along with Andreu Font-Ribera.  We extend our thanks to the referee for their constructive advise and valuable input.  Special thanks to Alan Aylward and Giovanna Tinetti for their support and advise over the course of this research.

\appendix

\section[]{Derivation of the model}

Here we derive the equations underlying the model presented in this paper.  First, consider an elliptical orbit in a frame $S$.  In this frame, the Cartesian co-ordinate of the planet is given by:

\begin{equation}
\frac{x^2}{a^2}+\frac{y^2}{b^2}=1
\end{equation}
\emph{where $a$ = semi-major axis and $b$ = semi-minor axis}

Now consider rotating the ellipse by an angle $\varpi$ about the z-axis in the counter-clockwise direction.  We now have the standard diagram for an orbit in top-view. The $S_1$ frame, in which the orbital inclination is $90^\circ$ and the observer and position of periastron are defined as is historically standard.  We also place the observer at infinite $x_1$ position.  The new axes are defined by:
\begin{equation}
x_1=x \cos \varpi-y \sin \varpi
\end{equation}
\begin{equation}
y_1=x \sin \varpi+y \cos \varpi
\end{equation}

Now we wish to find the projected orbital view, so we must change our axes a little.  The standard observer would see the star on their LHS for $\varpi$=0, rather than the RHS as presently defined, so let us change the x's and y's accordingly and define this frame $S_2$:
\begin{equation}
x_2=-x_1=-x \cos \varpi+y \sin \varpi
\end{equation}
\begin{equation}
y_2=-y_1=-x \sin \varpi-y \cos \varpi
\end{equation}

Now, let us now incline the orbit by the orbital inclination angle $i$, which is done by rotating the $S_2$ frame ellipse about the x-axis in a clockwise sense.  This defines the $S_3$ coordinates as:
\begin{equation}
x_3=x_2
\end{equation}
\begin{equation}
y_3=y_2 \cos i+z_2 \cos i=y_2 \cos i
\end{equation}
\begin{equation}
z_3=-y_2 \sin i+z_2 \cos i = -y_2 \sin i = -y_3 \tan i
\end{equation}

We substitute (A4) \& (A5) into (A6) \& (A7) and rearrange:
\begin{equation}
x=-y_3 \sec i \cdot \sin \varpi-x_3 \cos \varpi
\end{equation}
\begin{equation}
y=x_3 \sin \varpi-y_3 \sec i \cdot \cos \varpi
\end{equation}

We now substitute equations (A9) and (A10) into equation (A1) and solve for $y_3$.  The resulting analytic solution is the equation for the ellipse in transformed $S_3$ frame.
   
\begin{equation}
y_{3\alpha,3\beta}=\frac{\cos i [e x_3 \sin (2\varpi) \mp \sqrt{2 (1-e^2) (q a^2-2x_3^2)}]}{q} 
\end{equation}
\emph{where}
\begin{equation}
q=2-e^2+e^2 \cos (2\varpi)
\end{equation}
Now let us shift the origin to be centred about the star's position, the $S_4$=$S_{FINAL}$ frame.
\begin{equation}
x_4=x_3+a e \cos \varpi
\end{equation}
\begin{equation}
y_4=y_3+a e \sin \varpi \cdot \cos i
\end{equation}

Also note:
\begin{equation}
z_4=z_3+a e \sin \varpi \cdot \sin i
\end{equation}

By considering the points along the projected orbit which intersect a circle of radius $L$, in the $x_4$-$y_4$ plane, centred upon the starÕs position, we can determine the start \& end points for the primary \& secondary transits.  Hence, we simply substitute (A11) into equation (A14) and solve for $x_4$:

\begin{equation}
x_4^2+y_4^2=L^2
\end{equation}
\begin{equation}
x_4,solution=f(R_*,p,i,\varpi,a,e)
\end{equation}

The resulting analytic expression for $x_4$ is not concise but is possible to express with a number of substitutions, which we show below.  Also note that L may now be replaced with $L_B$ or $L_S$ as desired.

\begin{equation}
v = 1-e^2
\end{equation}

\begin{equation}
\alpha = 1+\frac{\cos^4 i(2e^2-4+q)^2+2\cos^2 i \Big[(2e^2+q)(4-q)-8\Big]}{q^2}
\end{equation}

\begin{align}
\varrho &= (4q^2)^{-1} \Big[-a^2 v^2 (2+13e^2)+L^2 (7e^4+8v)
\nonumber \\
&\qquad + a^2 v^2 \cos(4i) \cdot (8-6e^2-3q)+L^2 (e^4-8v) \cos(2i)\Big]
\end{align}

\begin{align}
\Upsilon &= (4q^2)^{-1} \Big\{-8 a^2 v^2 e^2 \cos(2i) (2 + \cos(2 \varpi))+ e^2 \Big[(-5a^2 v^2
\nonumber \\
&\qquad + 8L^2 (1+v) ) \cos(2\varpi)+2e^2 L^2 \cos(4\varpi) \sin^2i \Big] \Big\}
\end{align}

\begin{equation}
\beta=\Upsilon+\varrho
\end{equation}

\begin{equation}
\Delta=L^4+\frac{(2 a v L \cos i )^2 (q-4)+4(a v \cos i )^4}{q^2}
\end{equation}

\begin{equation}
\gamma=\frac{-2-3e^2+(2-e^2)\cos(2i)+2e^2 \cos(2\varpi) \sin^2(i)}{4}
\end{equation}

\begin{equation}
\epsilon=\frac{a^2 v^2 [1+\cos(2i)]+L^2 (q-4)}{2}
\end{equation}

\begin{equation}
\Lambda=-a e v \cos^2i \cos \varpi
\end{equation}

\begin{equation}
\kappa=\frac{(48 \Lambda)^2 \cdot (3 \gamma^2 \Delta - 2 \beta \gamma \epsilon + 3 \alpha \epsilon^2)}{q^4} - 16\beta (\beta^2-9 \Delta \alpha)
\end{equation}

\begin{equation}
\zeta = 4\beta^2+12\alpha \Delta + 768\epsilon \gamma q^{-4} \Lambda^2
\end{equation}

\begin{equation}
\Gamma=\frac{2 \cdot 2^{1/3} \cdot \zeta}{6 \alpha (\kappa+\sqrt{\kappa^2-4 \zeta^3})^{1/3}}+\frac{2^{2/3} \cdot (\kappa+\sqrt{\kappa^2-4 \zeta^3})^{1/3}}{6 \alpha}
\end{equation}

\begin{equation}
\Xi=\frac{64 \gamma^2 \Lambda^2}{q^4 \alpha^2}+\frac{4 \beta}{3 \alpha}
\end{equation}

Finally giving:

\begin{align}
x_4 &= -\frac{ 4 \gamma \Lambda }{ \alpha q^2 } + \frac{1}{2} \Bigg( \mp_b (\Gamma + \Xi)
\nonumber \\
&\qquad \mp_a \sqrt{ 2 \Xi - \Gamma \pm_b \frac{ 32 \Lambda \cdot (32 \alpha \gamma^3 \Lambda^2+\beta \gamma \alpha^2 q^4 -\epsilon \alpha^3 q^4)}{\alpha^4 q^6 \sqrt{\Gamma+\Xi}}} \Bigg)
\end{align}

The solved equation comes from solving a quartic equation and hence we produce four solutions, which correspond to the start \& end of the primary \& secondary transits.  We are presently only interested in the primary transit, which corresponds to the positive $z_4$ solutions.  These may be distinguished by using (A8) \& (A15).  Using the transform equations, we back-transform our $S_4$ frame solution into the $S$ frame to get $x(x_4)$ and $y(y_4)$.  We then use simple trigonometry to acquire the true anomaly:

\begin{equation}
f= \left\{
\begin{array}{rl}
\arctan \vartheta & \textrm{if } (x - a e) > 0 \textrm{ \& } y > 0\\
\pi + \arctan \vartheta & \textrm{if } (x - a e) < 0 \textrm{ \& } y \in \Re\\
2\pi + \arctan \vartheta & \textrm{if } (x - a e) > 0 \textrm{ \& } y < 0
\end{array} \right.
\end{equation}

\emph{where we define}
\begin{equation}
\vartheta = \frac{y}{x- a e}
\end{equation}

We now take advantage of the conservation of angular momentum and integrate (A34) between the appropriate limits.

\begin{equation}
\textrm{d}t = \frac{\mu r^2}{J} \textrm{d}f
\end{equation}

which we integrate and then finally write the duration between points a and b as:

\begin{equation}
t_a - t_b = \frac{\mu a^2}{J} [D(f_a) - D(f_b)]
\end{equation}

where we define $D(f)$ to be the \emph{duration function}:

\begin{equation}
D(f) = 2 \sqrt{1-e^2} \arctan\Big[ \sqrt{\frac{1-e}{1+e}} \tan \frac{f}{2} \Big]- \frac{e (1-e^2) \sin f }{1+e \cos f}
\end{equation}

We choose whether this is equivalent to $t_T$ or $t_F$ by selecting $L$ as being equal to $L_B$ or $L_S$ respectively, in the equations for $x_4$.  To employ our model with the Mandel-Agol limb darkening code, we simply produce the projected planet-star separation, $z$, and input it into the Mandel-Agol equations.  In our code, we produce a table of values of $z$ at different times, and then input the values into the Mandel-Agol model, in the usual way, to produce our lightcurves.

\begin{equation}
z = \frac{\sqrt{x_4^2 + y_4^2}}{R_*}
\end{equation}

We now show that by setting $e=0$, it can be easily shown that these equations analytically reduce down to the SMO equations for the timings.  The substitutions reduce down and we find the $x_4$ quartic solution reduces down to two solutions.

\begin{equation}
x_{4A}=-\csc i \sqrt{L^2-a^2 \cos^2i}
\end {equation}
\begin{equation}
x_{4B}=\csc i \sqrt{L^2-a^2 \cos^2i}
\end{equation}

Using equation (A16), we compute the $y_4$ solutions too.

\begin{equation}
y_{4A}=-\cos i \sqrt{(a^2-L^2) \csc^2i}
\end {equation}
\begin{equation}
y_{4B}=\cos i \sqrt{(a^2-L^2) \csc^2i}
\end{equation}

Note, that the sum of the squares of these two solutions is equal to $L^2$, as expected.  Converting these solutions back to the $S$ frame we find two solutions of positive $z_4$:

\begin{equation}
x_A=\csc i \cdot \sqrt{L^2-a^2 \cos^2i}
\end{equation}
\begin{equation}
x_B=-\csc i \cdot \sqrt{L^2-a^2 \cos^2i}
\end{equation}
\begin{equation}
y_A = y_B = -\sqrt{(a^2-L^2) \csc^2i}
\end{equation}

Note that once again the sum of the squares equals the expected result of $a^2$.
We therefore derive the total angular width of the transit to be:

\begin{equation}
f_{end}-f_{start}=2 \arcsin \Big( \frac{\sqrt{L^2-a^2 \cos^2i}}{a \sin i} \Big)
\end{equation}

Dividing this value by $2 \pi$ and multiplying by the period, $P$, gives the transit duration and we find that the result is the same as that derived by \citet{sea03} (specifically equations (2) \& (3) in the referenced paper by choosing $L=L_B$ or $L_S$ for $t_T$ and $t_F$ respectively).

\bsp

\label{lastpage}

\end{document}